# TlP$_5$: An unexplored direct band gap 2D semiconductor with ultra-high carrier mobility

*Jun-Hui Yuan,[1] Alessandro Cresti,[2] Kan-Hao Xue,[1,2,*] Ya-Qian Song,[1] Hai-Lei Su,[1] Li-Heng Li,[1] Nai-Hua Miao,[3,#] Zhi-Mei Sun,[3] Jia-Fu Wang,[4] Xiang-Shui Miao[1]*

[1]Wuhan National Research Center for Optoelectronics, School of Optical and Electronic Information, Huazhong University of Science and Technology, Wuhan 430074, China

[2]IMEP-LAHC, Grenoble INP – Minatec, 3 Parvis Louis Néel, 38016 Grenoble Cedex 1, France

[3]School of Materials Science and Engineering, Beihang University, Beijing 100191, China

[4]School of Science, Wuhan University of Technology, Wuhan 430070, China

## Corresponding Authors

[*]E-mail: xkh@hust.edu.cn (K.-H. Xue)

[#]E-mail: nhmiao@buaa.edu.cn (N.-H. Miao)

**ABSTRACT:** Two-dimensional materials with a proper band gap and high carrier mobility are urgently desired in the field of nanoelectronics. We propose a novel two-dimensional crystal monolayer TlP$_5$, which is dynamically and thermodynamically stable and possesses a direct band gap of 2.02 eV with high carrier mobilities (13960 cm$^2$ V$^{-1}$s$^{-1}$ for electrons and 7560 cm$^2$ V$^{-1}$s$^{-1}$ for holes), comparable to that of phosphorene. The band gap value and band characteristics of monolayer TlP$_5$ can be adjusted by biaxial and uniaxial strains, and excellent optical absorption over the visible-light range is predicted. These properties, especially for the balanced high mobilities for not only the electrons but also the holes, render monolayer TlP$_5$ an exciting functional material for future nanoelectronics and optoelectronic applications.

**KEYWORDS:** 2D materials, thallium penta-phosphorus, monolayer TlP$_5$, mobility, electronic properties, phosphorene, density functional theory

# 1 Introduction

Since the successful mechanical exfoliation of graphene in 2004,[1,2] two-dimensional (2D) materials such as silicene,[3] borophene,[4,5] phosphorene,[6,7] and transitional metal dichalcogenides (TMDCs)[8–11] have attracted intensive interests because of their extraordinary electrical, mechanical and thermal properties that enable extensive application potential in various fields.[9,12–18] For example, phosphorene, with a direct 2.0 eV band gap as well as a high carrier mobility of about $1.14 \times 10^3$ cm$^2$ V$^{-1}$ s$^{-1}$—$2.60 \times 10^4$ cm$^2$ V$^{-1}$ s$^{-1}$, has been considered to be a candidate for high performance field effect transistors.[6] Yet, its chemical instability under ambient conditions and the low electron mobility still hinder its practical application. MoS$_2$ possesses a suitable direct band gap (1.8 eV) for microelectronic and optoelectronic applications, but it suffers from the low carrier mobility (~200 cm$^2$ V$^{-1}$ s$^{-1}$). Nowadays, 2D materials with a proper band gap and high carrier mobilities for both electrons and holes are still urgently desired for logic and optoelectronic devices.

Very recently, a series of 2D phosphides, such as InP$_3$,[19] GeP$_3$,[20] SnP$_3$,[21] and CaP$_3$,[22] have been theoretically proposed as novel 2D semiconductors with high carrier mobilities (~10$^3$ cm$^2$ V$^{-1}$ s$^{-1}$—10$^4$ cm$^2$ V$^{-1}$ s$^{-1}$) that are comparable to those of phosphorene. In addition, their theoretical cleavage energy is relatively low (0.57~1.32 J m$^{-2}$), indicating that they can be obtained through mechanical exfoliation from bulk. On the other hand, a layered material composed of P and Tl with the TlP$_5$ stoichiometry was already reported in 1971 by Olofsson *et al*.[23] Experimentally, TlP$_5$ single crystal can be obtained under simple and easy synthetic conditions, which facilitates its mass production. However, there has been no relevant researches published on monolayer TlP$_5$ yet, and it is highly worthwhile to obtain a comprehensive understanding of the monolayer form.

In this work, we report a new monolayer phosphide, namely, 2D TlP$_5$ with a unique structure and high dynamic and thermal stability, using first-principles calculations. In addition, TlP$_5$ shows remarkably weak interlayer interactions, which result in a relatively low cleavage energy of 0.39 J m$^{-2}$. Our calculations suggest that TlP$_5$ monolayer is a semiconductor with 2.02 eV direct band gap. Moreover, it possesses remarkably high carrier mobilities of 13960 cm$^2$ V$^{-1}$ s$^{-1}$ and 7560 cm$^2$ V$^{-1}$ s$^{-1}$ for electrons and holes, respectively. It also shows strong light absorption in the visible-light and infrared regions, which is useful for optoelectronic devices.

## 2 Computational methods:

All density functional theory (DFT) calculations were performed using plane-wave-based Vienna *Ab-initio* Simulation Package (VASP).[24,25] The generalized gradient approximation (GGA) within the Perdew-Burke-Ernzerhof (PBE)[26] functional form was used for the exchange-correlation energy, and projector augmented-wave (PAW) pseudopotentials[27,28] were used to replace the core electrons. The HSE06 screened hybrid functional[35] was used to calculate the band structures in order to correct the band gaps in GGA-PBE. The plane wave energy cutoff was fixed to be 500 eV. The van der Waals (vdW) interactions were corrected by the DFT-D3 approach.[29] For all structural relaxations, the convergence criterion for total energy was set to $1.0\times10^{-6}$ eV, and structural optimization was obtained until the Hellmann-Feynman force acting on any atom was less than 0.01 eV/Å in each direction. The phonon dispersion was calculated with the density functional perturbation theory, using the PHONOPY code.[30] *Ab initio* molecular dynamics (AIMD) simulations were performed to examine the thermal stability of the structures, where NVT canonical ensembles were used.

## 3 Results and discussion

As shown in **Fig. 1(a)-1(c)**, the symmetry for bulk $TlP_5$ is orthorhombic with space group *Pmc*$2_1$, and the unit cell consists of four $TlP_5$ formula units. Our optimized lattice parameters of bulk $TlP_5$ are $a = 6.48$ Å, $b = 7.01$ Å and $c = 12.24$ Å, in good accordance with the experimental results ($a = 6.46$ Å, $b = 6.92$ Å and $c = 12.12$ Å).[23] In bulk $TlP_5$, the P atoms are connected to each other in a two dimensional network which is parallel to the (010) plane, while the two non-equivalent Tl atoms are bonded to their two neighboring P atoms in the same phosphorus layer. The phosphorus network existing in $TlP_5$ is very similar to that observed in the monoclinic modification of Hittorf's phosphorus[31] (see **Fig. S2** and **S3** for details). Thus, $TlP_5$ may be regarded as a particular 'phosphorene' whose surface has been passivated by metal thallium. The unique structural properties of $TlP_5$ may bring about distinct material characteristics. In addition, bulk $TlP_5$ is an indirect band gap semiconductor with a band gap value calculated to be 1.14 eV and 1.72 eV at the PBE and HSE06 levels (**Fig. 1(d)**), respectively.

Monolayer TlP$_5$ was obtained in our calculation by taking an atomic layer from bulk TlP$_5$ along the (010) direction. As shown in **Fig. 2(a)**, its structural characteristics remain, exhibiting a rectangular configuration. The optimized lattice parameters of monolayer TlP$_5$ are *a* =12.35 Å and *b* = 6.51 Å. The P-P bond length is 2.16 Å to 2.24 Å, shorter than that of phosphorene (2.24 Å to 2.28 Å), while the Tl-P bond length is 3.00 Å to 3.17 Å (see **Table S1**). As is well known, mechanical cleavage[1,6] and liquid phase exfoliation [32] are powerful techniques to produce single and few layer flakes from the layered bulk materials. To assess such possibility we have calculated the cleavage energy of monolayer TlP$_5$ from a five-layer TlP$_5$ slab, mimicking the bulk. As shown in **Fig. 2(b)**, the cleavage energy increases with the interlayer distance, but reaching a convergence of about 0.39 J m$^{-2}$. The estimated exfoliation energies of graphene and black phosphorus are, respectively, 0.32 J m$^{-2}$ and 0.37 J m$^{-2}$ according to our calculation, which are in line with previous theoretical studies.[19,33] The DFT-estimated exfoliation energies for some other layered phosphides such as InP$_3$,[19] GeP$_3$,[20] CaP$_3$[22] and SnP$_3$[21] are 1.32 J m$^{-2}$, 1.14 J m$^{-2}$, 1.30 J m$^{-2}$ and 0.57 J m$^{-2}$, respectively. Therefore, exfoliation from the bulk is feasible for monolayer TlP$_5$ preparation, as its cleavage energy is comparable with the mentioned 2D materials above.

In addition, the phonon dispersions of monolayer TlP$_5$ (shown in **Fig. 2(c)**) consist of only real modes whose styles are typical for 2D crystals, indicating the kinetic stability. The highest frequency mode of monolayer TlP$_5$ reaches 477 cm$^{-1}$, which is very close to that of MoS$_2$ (473 cm$^{-1}$)[34] and higher than that of phosphorene (~450 cm$^{-}$ [1])[16], revealing the mechanical robustness of the covalent P-P bonds. The thermal stability is further substantiated by *ab initio* molecular dynamics (AIMD) simulations (see **Fig. S5**), where the monolayer $TlP_5$ structure remains intact at 300 K after a 5 *ps* simulation time.

Subsequently, we studied the electronic band structures of monolayer $TlP_5$. Since Tl is a heavy element, the spin-orbit coupling (SOC) effect may affect the electronic properties of monolayer $TlP_5$. The band structures of monolayer $TlP_5$ with and without SOC were both calculated, using either the PBE functional or the HSE06 hybrid functional. As shown in **Fig. 3(a)**, using GGA-PBE without considering SOC, the band gap of $TlP_5$ is predicted to be a 1.37 eV direct one. A similar band structure but with a larger direct band gap of 2.02 eV has been confirmed based on HSE06 calculations. We then examine the effect of SOC in our calculations. The conduction band minimum (CBM) and valence band maximum (VBM) of monolayer TlP5 do not show discernable shift after turning on the SOC, regardless of using PBE or HSE06. For the band gap, the difference is less than 0.01 eV with and without SOC (**Fig. S6a** and **S6b**). The negligible impact of SOC indicate that it is the P $3p$ orbitals that contribute most to the VBM and CBM states, as can be confirmed by the wave functions and partial density of states (PDOS) results (shown in **Fig. S7a** and **S7b**). Therefore, we shall neglect the SOC effect in all forthcoming band structure calculations.

In order to elucidate the changes in the electronic properties of $TlP_5$ from the bulk to few layers, we have investigated the electronic band gaps of 2D $TlP_5$ with varying number of layers, where the results are shown in **Fig. 3(b)** and **Fig. S8**. The electronic structures of 2D TlP$_5$ multilayers indeed strongly depend on the number of layers. The bi-layer and tri-layer TlP$_5$ still maintain the direct band gap feature, similar to the monolayer TlP$_5$. However, the four-layer TlP$_5$ (**Fig. S8c**) shows a direct-to-indirect band gap transition with the VBM shifted from $Y$ to $A$ (which lies along the $Y$-$\Gamma$ direction), though the CBM remains at the $Y$-point. The five-layer TlP$_5$ remains as an indirect band gap semiconductor (**Fig. S8d**). The possible reasons for losing the direct gap feature are the lacking of interlayer interaction, and structural reconstruction upon stacking atomic layers.[22]

In order to explore the application potential of monolayer TlP$_5$ in electronic devices, we systematically calculated the carrier mobility (electrons and holes) based on the deformation potential (DP) theory proposed by Bardeen and Shockley.[35] The carrier mobility of 2D materials can be evaluated by the following equation[22,36,37]

$\mu_{2D} = \dfrac{e\hbar^3 C_{2D}}{k_B T m^* m_d (E_1^i)^2}$, where $\hbar$ is the reduced Planck constant, $k_B$ is the Boltzmann constant, $m^*$ is the effective mass in the direction of transport, $m_d$ is the average effective mass determined by $m_d = (m_a^* m_b^*)^{1/2}$, and $T$ is the temperature ($T$ = 300 K). The elastic modulus $C_{2D}$ of the longitudinal strain in the propagation direction is derived from $(E - E_0)/S_0 = C_{2D}(\Delta l / l_0)^2 / 2$, where $E$ is the total energy of the 2D structure, and $S_0$ is the lattice area of the equilibrium supercell. The deformation potential constant $E_1^i$ is defined as $E_1^i = \Delta E_i / (\Delta l / l_0)$. Here $\Delta E_i$ is the energy change of the $i^{th}$ band under proper cell compression and dilatation (calculated using a step of 0.25%), $l_0$ is the lattice constant in the transport direction and $\Delta l$ is the deformation of $l_0$.

As summarized in **Table 1**, the elastic moduli are obviously anisotropic with the value of 21.59 N m$^{-1}$ and 88.49 N m$^{-1}$ along the *a* and *b* directions, respectively. These are slightly lower than those of phosphorene ($C_x$ = 29 N m$^{-1}$ and $C_y$ =102 N m$^{-1}$)[36]. The effective masses of both electron and hole along the *a*/*b* directions are anisotropic as well. Especially for the hole, the effective mass along *a* direction (0.886 $m_e$) is notably larger than that along the *b* direction (0.215 $m_e$). The deformation potential $E_l$ is nearly the same for the electron (0.574 eV along *a* and 0.614 eV along *b*), but exhibits a strong anisotropy for the hole (similar to hittorfene[31], phosphorene[36] and CaP$_3$[22]) with that of the *b* direction having the smallest value of about 0.397 eV. This contributes to a very large hole mobility along the *a* direction, up to 7.56×10$^3$ cm$^2$ V$^{-1}$ s$^{-1}$. Note that the electron mobility along *b* direction points to an even larger value of 1.396×10$^4$ cm$^2$ V$^{-1}$ s$^{-1}$, which is because of the joint action of a large elastic modulus and a small deformation potential along the *b* direction. It is also noteworthy that even the lower electron mobility in *a* direction (5.24×10$^3$ cm$^2$ V$^{-1}$ s$^{-1}$) and hole mobility in *b* direction (1.51×10$^3$ cm$^2$ V$^{-1}$ s$^{-1}$) are of similar order of magnitude compared with other 2D phosphides, while much larger than that of most 2D TMDCs[38].

Compared with phosphorene and their derivatives reported recently, monolayer TlP$_5$ possesses one of the most attractive electronic properties (see **Table S3**). Traditionally, phosphorene and their derivatives have either high hole mobility combined with low electron mobility, or vice versa. For instance, monolayer CaP$_3$ has a high electron mobility of 1.99×10$^4$ cm$^2$ V$^{-1}$ s$^{-1}$, whereas its hole mobility is only 0.78 ×10$^3$ cm$^2$ V$^{-1}$ s$^{-1}$.[22] However, high hole and electron mobilities of nearly or more than the order of 10$^4$ cm$^2$ V$^{-1}$ s$^{-1}$ can be achieved simultaneously in monolayer TlP$_5$, which is of great significance for improved device performance. Furthermore, the structure of 2D TlP$_5$ is different from the reported 2D InP$_3$ or CaP$_3$.[19,22] There is a continuous framework of P in TlP$_5$, involving two different P clusters, *i.e.* P$_6$ and P$_5$. Besides, anisotropic structures also lead to anisotropic electronic structures and mechanical properties as already discussed. Note that the monolayer TlP$_5$ possesses a direct band gap of 2.02 eV (calculated using HSE06, where the same criteria follow below), much higher than that of InP$_3$ (indirect, 1.14 eV), GeP$_3$ (indirect, 0.55 eV), CaP$_3$ (direct, 1.15 eV), SnP$_3$ (indirect, 0.72 eV) and phosphorene (direct, 1.51 eV). Therefore, monolayer TlP$_5$ would be particularly suitable for nanoelectronic and optoelectronic applications.

Since applying elastic strain is an effective means of band structure engineering in 2D semiconductors,[14,39] we further studied the effects of in-plane compressive/tensile biaxial and uniaxial strains on the band structures of monolayer TlP$_5$. **Figure 4** presents the HSE06-predicted band gaps of TlP$_5$ monolayer under different strains in the range of -6% to 6% (the band structures are plotted in **Fig. S10**). Interestingly, the band gaps of TlP$_5$ monolayer increase gradually with either compressive or tensile uniaxial strains along *a* axis, and the direct band gap feature remains unchanged within the range of strain being considered (see **Fig. S10a**). On the other hand, under the biaxial strain and the uniaxial strain along *b* axis, the band gap value demonstrates a similar variation trend, *i.e.*, increasing linearly until up to 2%, and then both decreasing linearly. A direct-to-indirect band gap transition has been discovered when the tensile strain reaches 3% for both biaxial and uniaxial strains along *b* axis. The position of the VBM is located at the *Y*-point while the CBM has shifted from *Y* to *Γ* (see **Fig. S10b** and **S10c**). The indirect band gap feature is maintained with the continuous increase of strain. The variable responses to the strain in monolayer TlP$_5$ may be of interest for certain mechanical applications.

The direct band gap feature of monolayer TlP$_5$ allows it to be directly coupled to light, like phosphorene.[6] Hence, we further explored the optical properties of TlP$_5$ monolayer by calculating the absorption spectra in- and out-of-plane using the HSE06 functional. The transverse dielectric function $\varepsilon(\omega) = \varepsilon_1(\omega) + i\varepsilon_2(\omega)$ is used to describe the optical properties of materials,[40] where $\omega$ is the photon frequency, $\varepsilon_1(\omega)$ is the real part and $\varepsilon_2(\omega)$ is the imaginary part of the dielectric function, respectively. The absorption coefficient can be evaluated according to the expression[40] $\alpha(\omega) = \frac{\sqrt{2}\omega}{c}\left\{\left[\varepsilon_1^2(\omega) + \varepsilon_2^2(\omega)\right]^{\frac{1}{2}} - \varepsilon_1(\omega)\right\}^{\frac{1}{2}}$. As shown in **Fig. 5**, the absorption coefficients of monolayer TlP$_5$ reaches the order of $10^5$ cm$^{-1}$, and covering a wide wavelength range in the visible light region. In-plane absorption is always stronger than that of out-of-plane, due to the larger cross section area. In addition, in-plane optical absorption shows obvious anisotropy along the *x* and *y* directions, as shown in **Fig. 5(a)** and **5(b)**. Interestingly, under the applied biaxial compressive strain, both the in-plane and out-plane absorption coefficients are greatly enhanced in the all photon-energy range, which may be attributed to the band gap reduction caused by the strain (see **Fig. S10c**). However, although the band gap decreases under the biaxial tensile strain as well, a slight reduction in optical absorption has been observed, which may stem from the direct-indirect band gap transition under strain (**Fig. S10c**). The outstanding optical properties suggest potential applications of monolayer TlP$_5$ as efficient optical absorber materials in solar cells and optoelectronic devices.

## 4 Conclusions

In summary, we have shown that the 2D semiconducting monolayer TlP$_5$ is a remarkable new candidate for high performance nanoelectronics and optoelectronic devices. Synthesis of the layered bulk semiconductor species has been known since 1971, and the predicted cleavage energies indicate that exfoliation from the bulk is possible. Monolayer TlP$_5$ shows a direct band gap of 2.02 eV with a quite balanced and outstanding carrier mobilities for electrons ($1.396\times10^4$ cm$^2$ V$^{-1}$ s$^{-1}$) and holes ($0.756\times10^4$ cm$^2$ V$^{-1}$ s$^{-1}$), even superior to that of phosphorene. Energy gap characteristics can be modulated for the monolayer by strain engineering. Besides, TlP$_5$ monolayer has a substantial light absorption in the range of the solar spectrum.

## Conflicts of interest

There are no conflicts to declare.

## Acknowledgements

This work was supported by the National Key Research and Development Program of China (Materials Genome Initiative, 2017YFB0701700), the National Natural Science Foundation of China under Grant No. 11704134, the Fundamental Research Funds of Wuhan City under Grant No. 2017010201010106, and the Fundamental Research Funds for the Central Universities of China under Grant No. HUST:2016YXMS212. K.-H. Xue received support from China Scholarship Council (No. 201806165012).

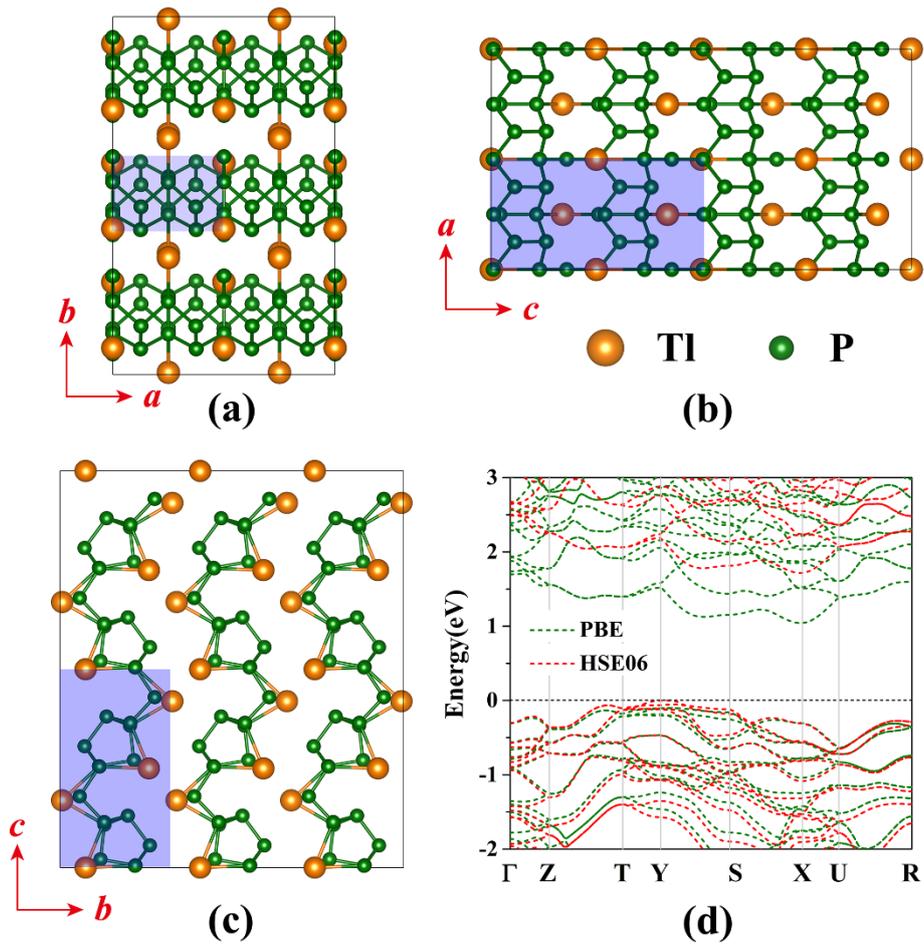

**Figure 1** (a) Top and side views (along $c$ and $b$ directions) of optimized bulk TlP$_5$ with 2×3×2 supercell, respectively. The unit cell is marked in light blue squares. (d) Calculated electronic band structures of bulk TlP$_5$ using GGA-PBE and HSE06, respectively. The Fermi levels are set to zero energy.

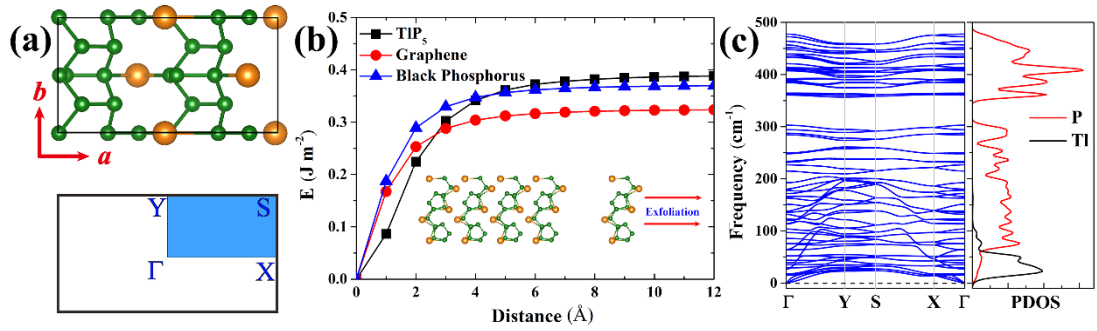

**Figure 2** (a) Top view of the optimized monolayer TlP$_5$ and the corresponding first Brillouin zone with high symmetry points. (b) Cleavage energy estimation for the formation of monolayer TlP$_5$, calculated by enlarging the interlayer distance between the monolayer system that is removed from the remainder of a five-layer slab, resembling the bulk model. (c) The calculated phonon dispersion spectra and phonon density of states of the monolayer TlP$_5$.

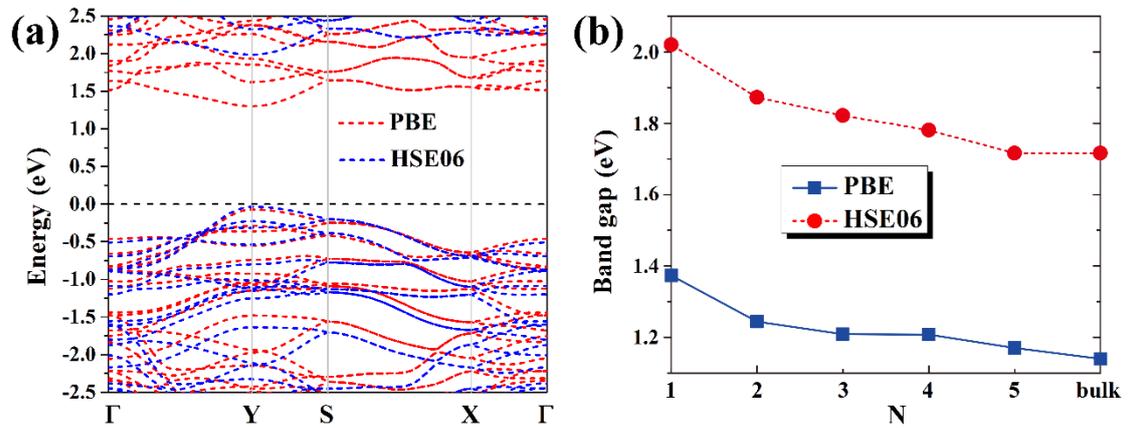

**Figure 3** (a) Electronic band structures of monolayer TlP$_5$ calculated using GGA-PBE and HSE06 without considering SOC. (b) Computed band gaps of TlP$_5$ multilayers versus the number of atomic layers, using GGA-PBE and hybrid HSE06 functionals, respectively.

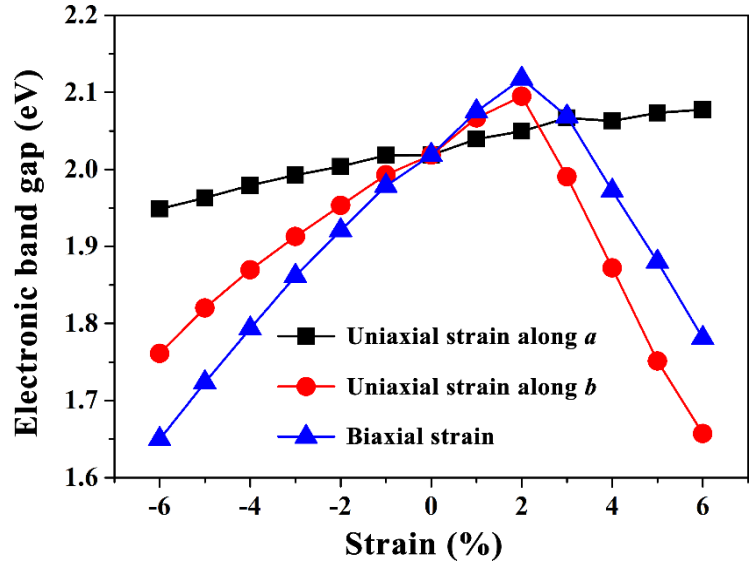

**Figure 4** Electronic band gaps of monolayer TlP$_5$ under various strains, calculated using screened HSE06 hybrid functional.

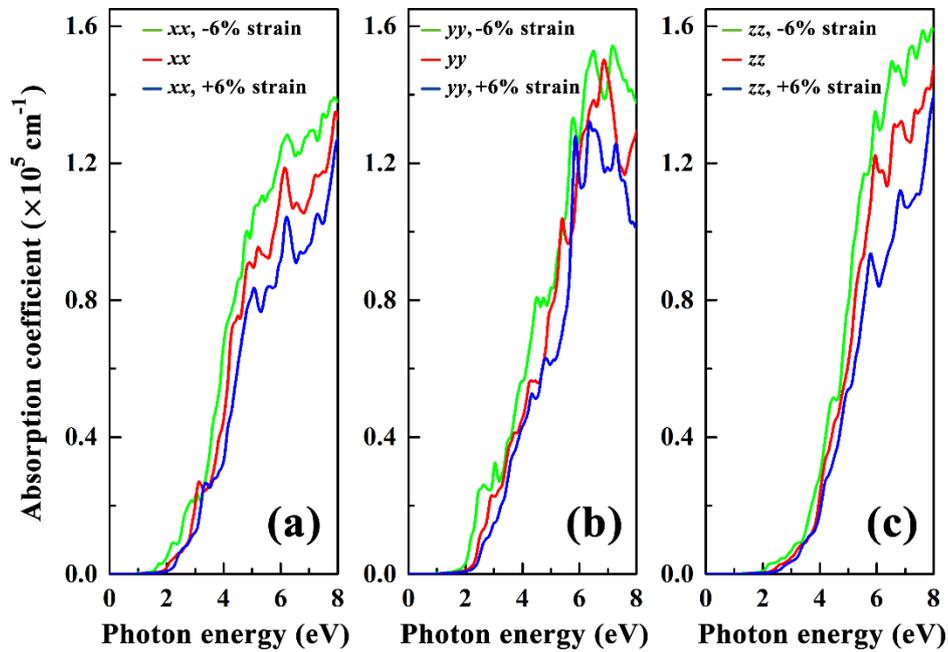

**Figure 5** Calculated in-plane ((a) and (b)) and out-plane (c) light absorption coefficients of monolayer TlP$_5$, using the screened HSE06 hybrid functional.

**Table 1** Calculated effective mass $m^*$ (unit: $m_e$), deformation potential constant $|E_1^i|$ (unit: eV), elastic modulus $C_{2D}$ (unit: N m$^{-1}$), and carrier mobility $\mu_{2D}$ (unit: $10^3$ cm$^2$ V$^{-1}$s$^{-1}$) for TlP$_5$ monolayer along the $a$ ($Y$-$S$) and $b$ ($Y$-$\Gamma$) directions.

| Carrier type | $m_a^*$ | $m_b^*$ | $|E_{1a}|$ | $|E_{1b}|$ | $C_a^{2D}$ | $C_b^{2D}$ | $\mu_a^{2D}$ | $\mu_b^{2D}$ |
|---|---|---|---|---|---|---|---|---|
| electrons | 0.480 | 0.645 | 0.574 | 0.614 | 21.59 | 88.49 | 5.24 | 13.96 |
| holes | 0.886 | 0.215 | 0.397 | 3.647 | 21.59 | 88.49 | 7.56 | 1.51 |

## TOC graphic

## TlP$_5$: An unexplored direct band gap 2D semiconductor with ultra-high carrier mobility

Jun-Hui Yuan,[1] Alessandro Cresti,[2] Kan-Hao Xue,[1,2]* Ya-Qian Song,[1] Hai-Lei Su,[1] Li-Heng Li,[1] Nai-Hua Miao,[3#] Zhi-Mei Sun,[3] Jia-Fu Wang,[4] Xiang-Shui Miao[1]

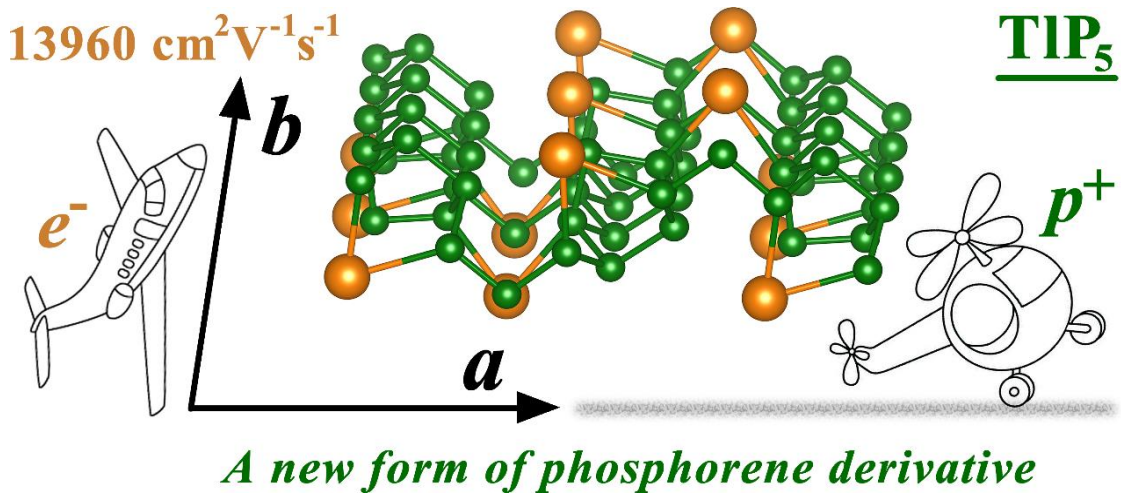

*A new form of phosphorene derivative*

# TlP$_5$: An unexplored direct band gap 2D semiconductor with ultra-high carrier mobility

Jun-Hui Yuan,[1] Alessandro Cresti,[2] Kan-Hao Xue,[1,2,*] Ya-Qian Song,[1] Hai-Lei Su,[1] Li-Heng Li,[1] Nai-Hua Miao,[3,#] Zhi-Mei Sun,[3] Jia-Fu Wang,[4] Xiang-Shui Miao[1]

[1] Wuhan National Research Center for Optoelectronics, School of Optical and Electronic Information, Huazhong University of Science and Technology, Wuhan 430074, China

[2] IMEP-LAHC, Grenoble INP – Minatec, 3 Parvis Louis Néel, 38016 Grenoble Cedex 1, France

[3] School of Materials Science and Engineering, Beihang University, Beijing 100191, China

[4] School of Science, Wuhan University of Technology, Wuhan 430070, China

*Correspondence and requests for materials should be addressed to K.-H. Xue and N.-H. Miao (email: xkh@hust.edu.cn; nhmiao@buaa.edu.cn)

First of all, we have checked the stability of bulk TlP$_5$ through the phonon dispersion calculation. The result in **Figure S1** shows only real modes, indicating good dynamic stability.

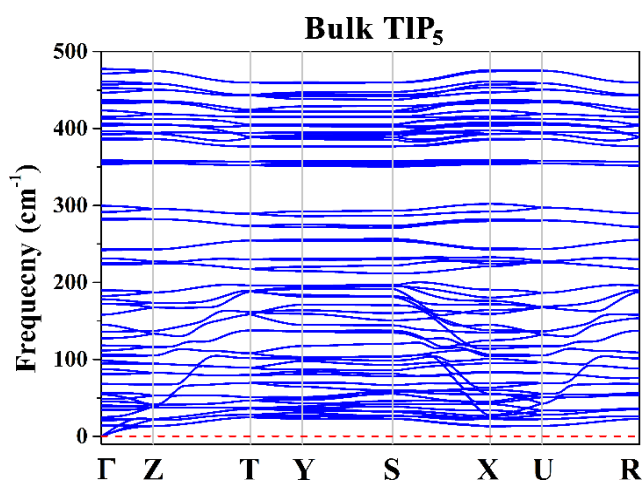

**Figure S1** Calculated phonon dispersion relation of bulk TlP$_5$.

The phosphorus network founded in TlP$_5$ is very similar to the one observed in the monoclinic modification of red phosphorus, the so-called Hittorf's phosphorus.[1] As shown in **Figure S2**, in TlP$_5$ there are discernable tubes with pentagonal cross-sections, which are very characteristic structure of the phosphorus element in Hittorf's phosphorus. Comparing these two structures, one finds that the phosphorus network in TlP$_5$ can be regarded as stemming from a partial breaking of the network in Hittorf's phosphorus, as shown in **Figure S3.**

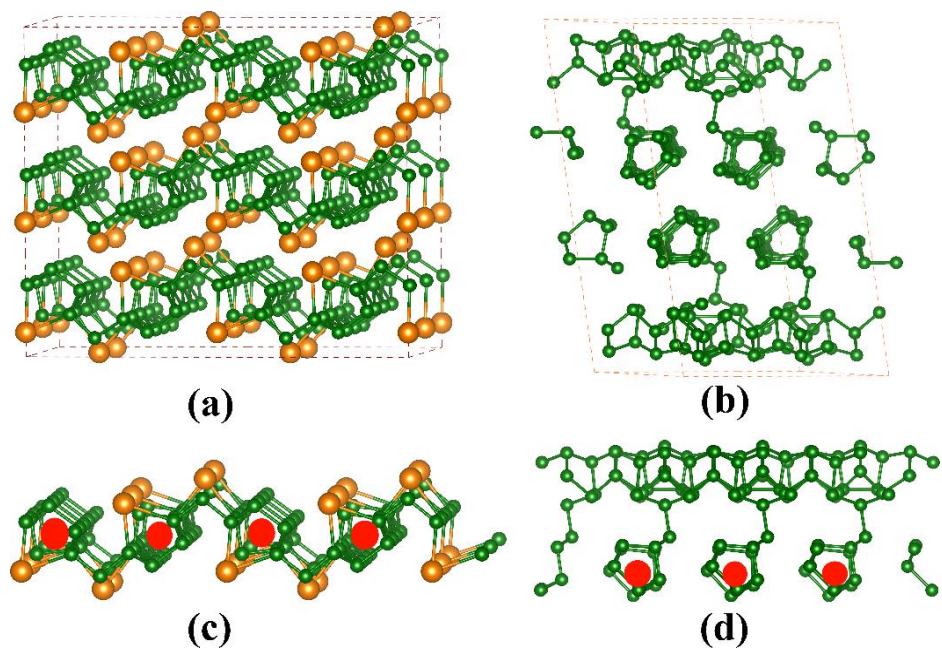

**Figure S2** Demonstration of the crystal structures for: (a) bulk (2×3×2 supercell) TlP$_5$; (b) bulk (1×2×1 supercell) Hittorf's phosphorus; (c) monolayer (2×2×1 supercell) TlP$_5$; and (d) monolayer (2×2×1 supercell) Hittorf's phosphorus.

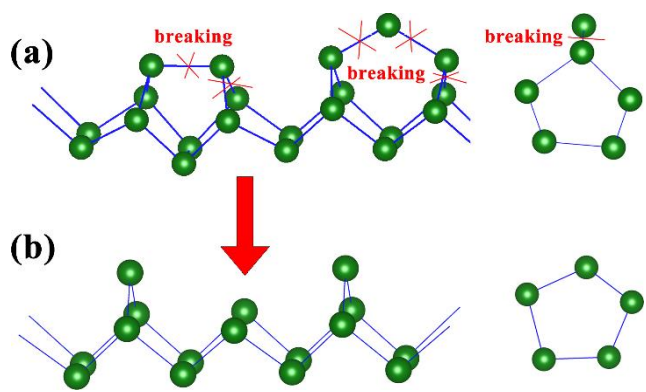

**Figure S3** Comparison of the pentagonal tubes in the phosphorus network of (a)

Hittorf's phosphorus and (b) TlP$_5$. Illustrations of the possible P-P bond breaking schemes are also given.

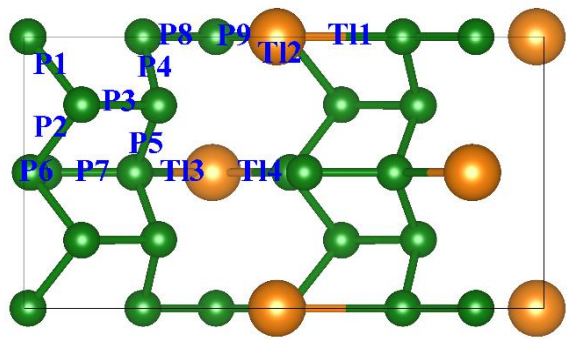

**Figure S4** A schematic diagram of the P-P bonds and Tl-P bonds in monolayer and bulk TlP$_5$. Only a single formula unit of TlP$_5$ has been marked.

The bond lengths of P-P and Tl-P in both monolayer and bulk TlP$_5$ are listed in Table S1, together with the experimental results for the sake of comparison. The P-P bond lengths of monolayer TlP$_5$ are around 2.16 Å ~2.27 Å, while those of Tl-P are around 3.0 Å ~3.17 Å. These values are very close to those of optimized bulk TlP$_5$ (2.15 Å~2.28 Å for P-P bonds and 3.00 Å~3.09 Å for Tl-P bonds). Most of the P-P and Tl-P bond lengths in monolayer TlP$_5$ are slightly larger than in the bulk, which is as expected.

**Table S1** The optimized P-P and Tl-P bond lengths in the TlP$_5$ monolayer ("1L" for short) and bulk, respectively, in comparison to the experimentally verified values (Exp.) for the bulk. The numbers in parentheses count the equivalent P-P bonds in a unit cell, and the corresponding bond locations are marked in **Figure S4**.

|       | 1L TlP$_5$ | Bulk TlP$_5$ | Exp.    |
|-------|-----------|--------------|---------|
|       | P-P (Å)   | P-P (Å)      | P-P (Å) |
| P1(4) | 2.243     | 2.242        | 2.242   |
| P2(4) | 2.218     | 2.216        | 2.213   |
| P3(4) | 2.27      | 2.275        | 2.229   |
| P4(4) | 2.238     | 2.228        | 2.222   |
| P5(4) | 2.227 | 2.218 | 2.225 |
| P6(2) | 2.174 | 2.172 | 2.174 |
| P7(2) | 2.157 | 2.15 | 2.126 |
| P8(4) | 2.187 | 2.178 | 2.13 |
| P9(2) | 2.203 | 2.198 | 2.221 |
| | P-Tl(Å) | P-Tl(Å) | P-Tl(Å) |
| Tl1(4) | 3.17 | 3.03 | 2.985 |
| Tl2(4) | 3.01 | 3.023 | 3.015 |
| Tl3(2) | 3.008 | 2.998 | 2.986 |
| Tl4(2) | 3.04 | 3.089 | 3.025 |

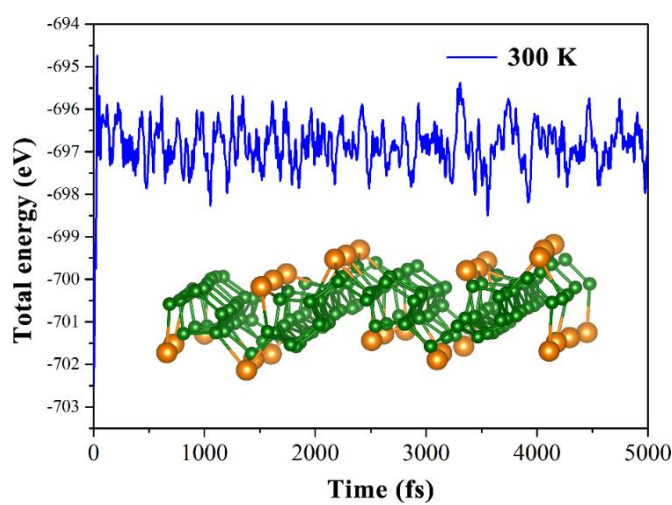

**Figure S5** Side view of the snapshots from the molecules dynamics simulation for the TlP$_5$ monolayer. The variation of total energy was recorded during the simulation time of 5 ps, and the temperature was 300 K.

The optimized lattice constants of $n$L ($n$=1, 2, 3, 4, 5) TlP$_5$ and bulk are listed in Table S2, with reference to experimental values. The optimized lattice parameters of monolayer TlP$_5$ are $a$ =12.35 Å and $b$ = 6.51 Å (note that the $a/b$ here actually correspond to the $c/a$ directions in bulk). Upon the increase in the number of layers, the lattice constants vary very little, indicating a relatively weak interlayer coupling in the TlP$_5$ system.

**Table S2** The optimized lattice constants (*a*/*b*/*c*) in TlP$_5$ monolayer (1L), bilayer (2L), tri-layer (3L), four-layer (4L), five-layer (5L) and bulk, respectively, in comparison to the experimental values (Exp.) for the bulk.

| Lattice constant | 1L | 2L | 3L | 4L | 5L | Bulk | Exp.[2] |
|---|---|---|---|---|---|---|---|
| *a*/Å | 12.35 | 12.27 | 12.26 | 12.26 | 12.25 | 6.48 | 6.46 |
| *b*/Å | 6.51 | 6.49 | 6.49 | 6.48 | 6.48 | 7.01 | 6.92 |
| *c*/Å | -- | -- | -- | -- | -- | 12.24 | 12.12 |

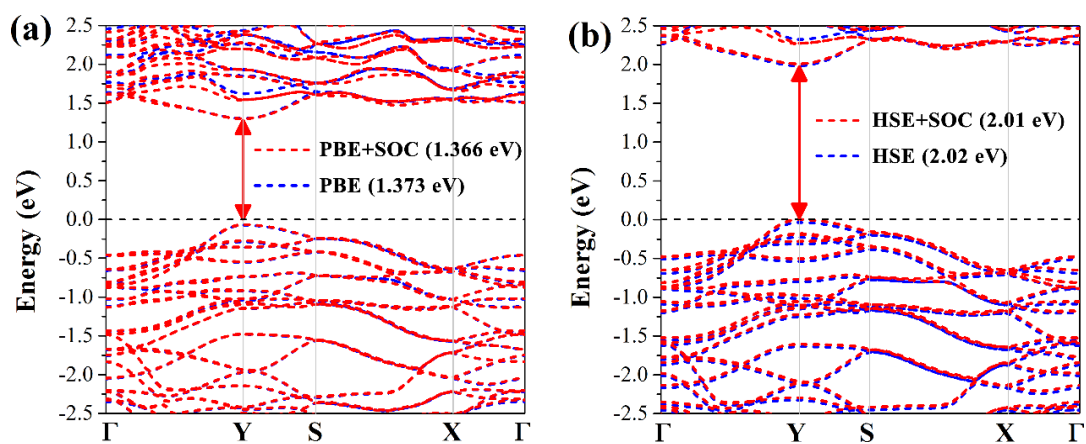

**Figure S6** (a) Band structures of monolayer TlP$_5$ calculated using the PBE functional, either with or without considering the effect of spin orbit coupling (SOC). (b) Band structures of monolayer TlP$_5$ calculated using the screened HSE06 hybrid functional, either with or without considering the effect of SOC.

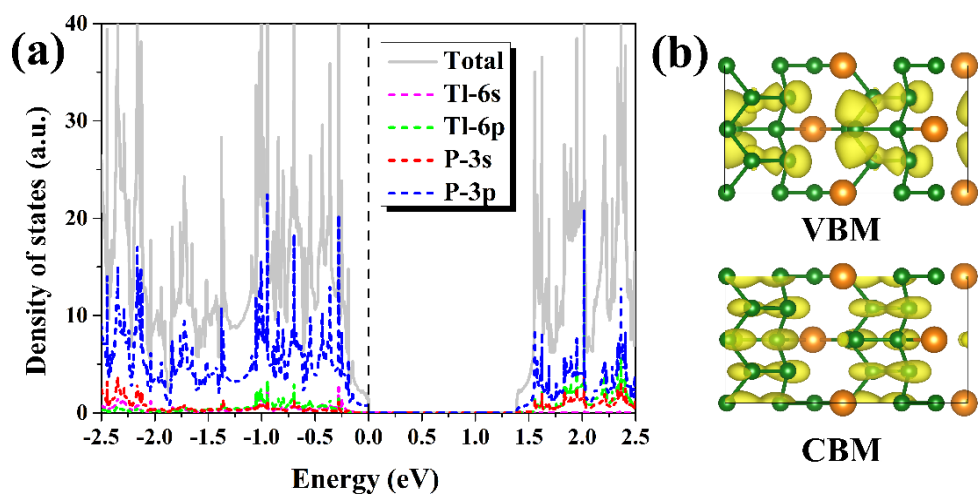

**Figure S7** (a) Projected density of states of monolayer TlP$_5$. (b) Isosurfaces of partial

S5

charge densities corresponding to the VBM and CBM of monolayer TlP$_5$.

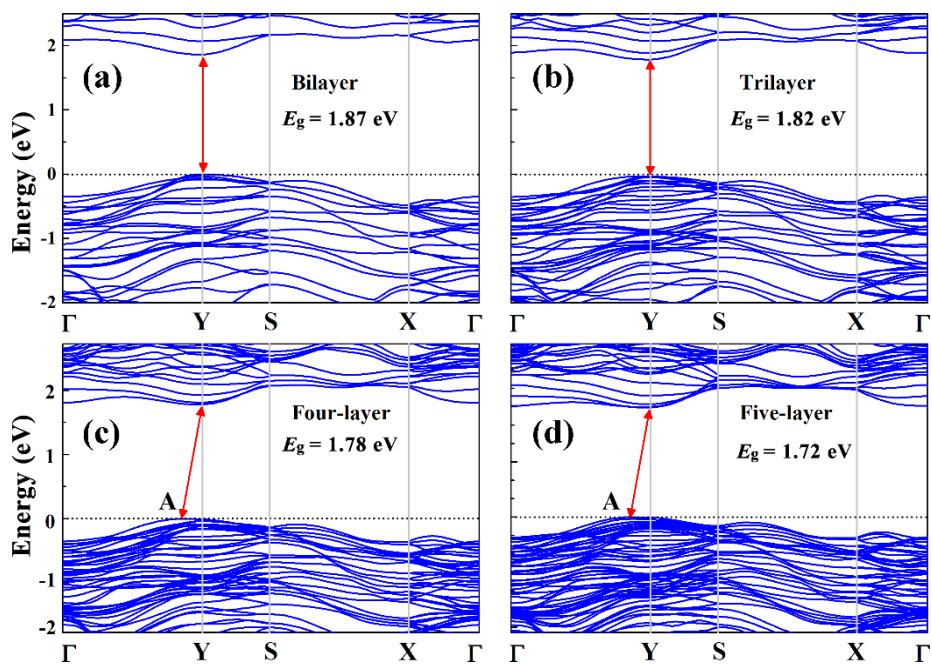

**Figure S8** Electronic band structures of 2D TlP$_5$ with varying number of layers, calculated using the screened HSE06 hybrid functional: (a) bilayer; (b) trilayer; (c) four-layer; and (d) five-layer.

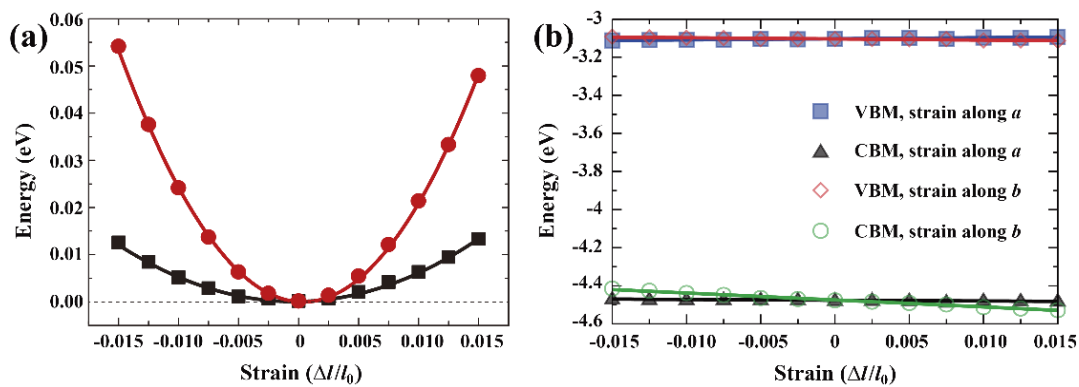

**Figure S9** (a) The relationship between total energy and the applied strain $\delta$ along the *a* and *b* directions of monolayer TlP$_5$. The quadratic data fitting gives the in-plane stiffness of 2D structures. Black and red curves show the in-plane stiffness along the *a* and *b* directions of monolayer TlP$_5$, respectively. (b) The shift of VBMs and CBMs for monolayer TlP$_5$ with respect to the vacuum energy, as a function of the applied strain along either the *a* or the *b* direction. The linear fit of the data yields the deformation potential constant.

**Table S3** The carrier mobilities $\mu_{2D}$ ($\times 10^3$ cm$^2$ V$^{-1}$s$^{-1}$) and energy bang gaps (eV, calculated at the HSE06 level) of phosphorene and their derivatives for comparison (only the monolayer is considered). The symbols *d* and *i* represent direct and indirect band gaps, respectively.

|  | Direction | TlP$_5$ | Phosphorene[3] | Hittorfene[1] | InP$_3$[4] | GeP$_3$[5] | CaP$_3$[6] | SnP$_3$[7] |
|---|---|---|---|---|---|---|---|---|
| Electron | x | 5.24 | 1.10~1.14 | 0.50 | 0.54 | 0.04 | 19.9 | 0.19 |
|  | y | 13.96 | 0.08 | 0.43 | 1.92 | 0.07 | 1.75 | 0.21 |
| Hole | x | 7.56 | 0.64~0.70 | 0.31 | 0.006 | 0.014(0.35) | 0.08 | 0.17 |
|  | y | 1.51 | 10~26 | 7.68 | 0.05 | 0.19(2.36) | 0.78 | 0.36 |
| Band gap |  | 2.02(*d*) | 1.51(*d*) | 2.5(*d*) | 1.14(*i*) | 0.55(*i*) | 1.15(*d*) | 0.72(*i*) |

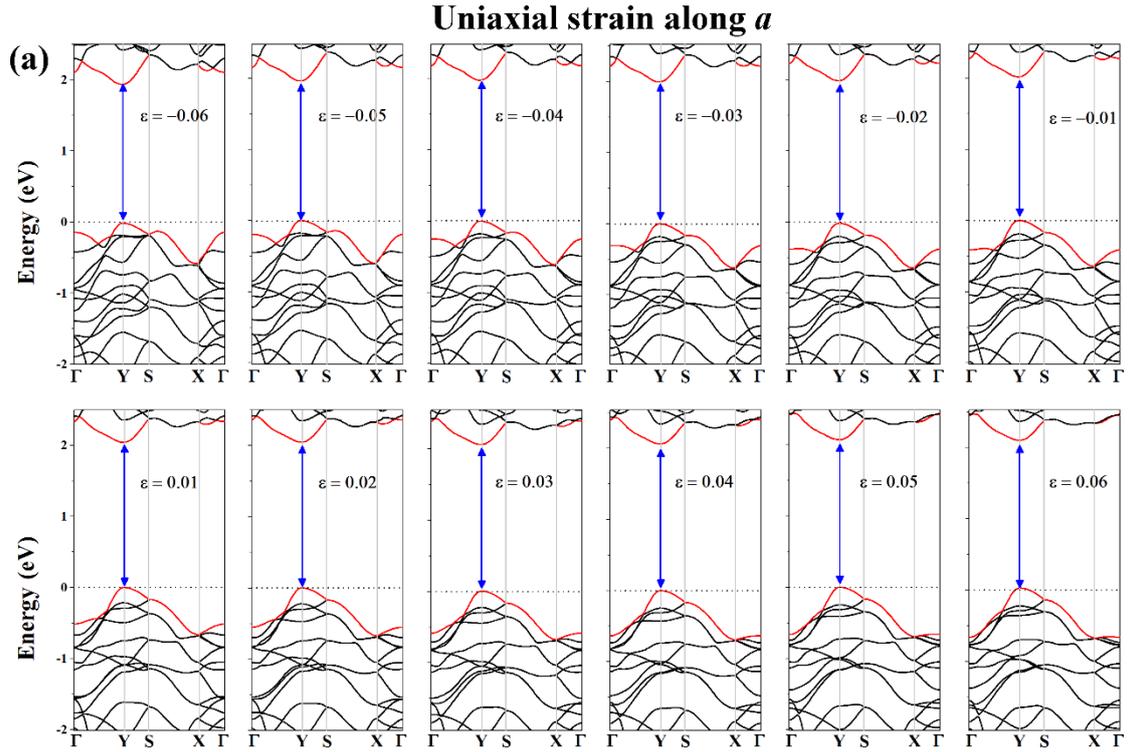

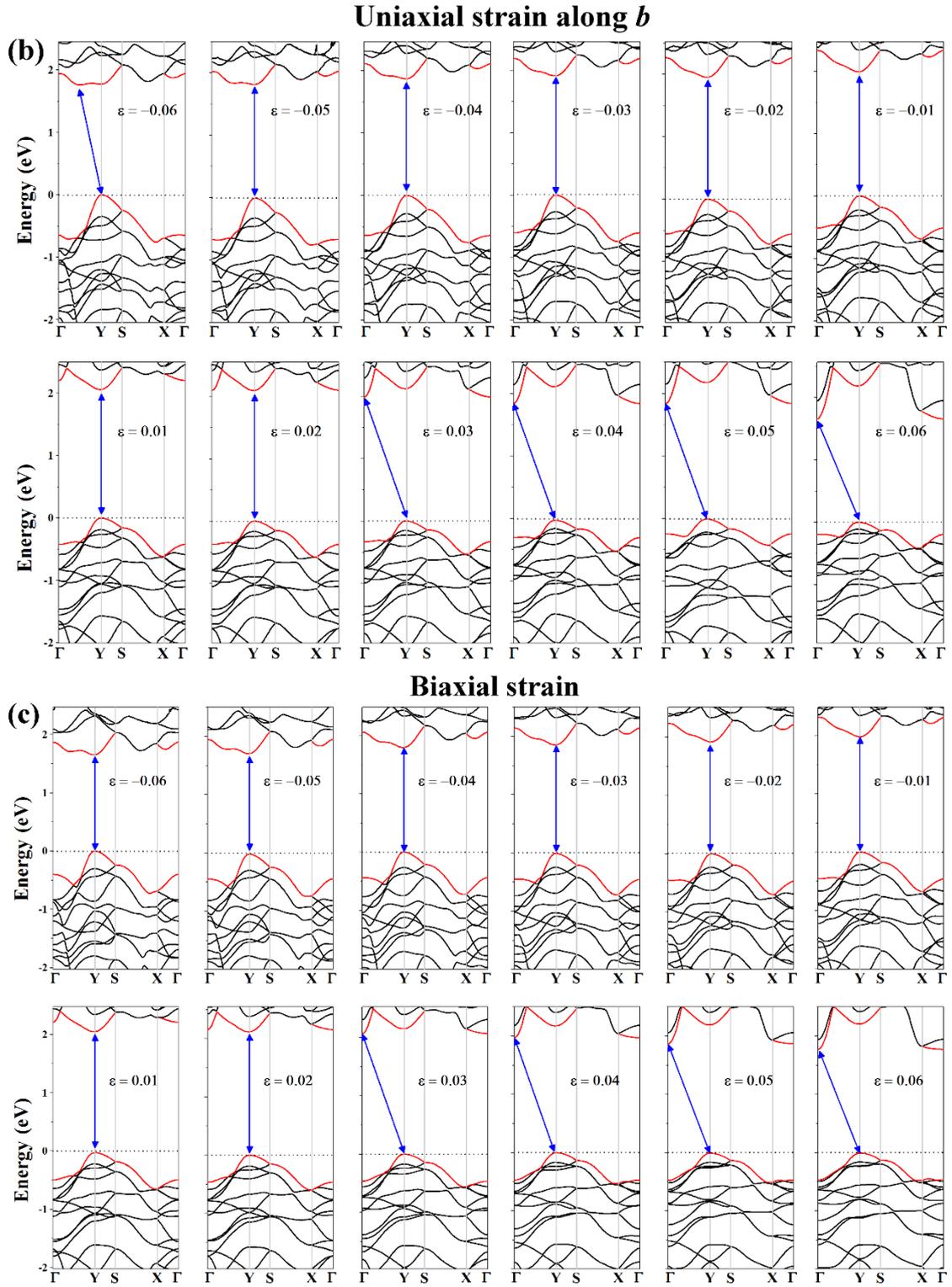

**Figure S10** Electronic band structures of monolayer TlP$_5$ under various strain situations, calculated using the HSE06 functional. The applied strains are (a) uniaxial strain along *a*-axis; (b) uniaxial strain along *b*-axis (b); and (c) biaxial along both *a* and *b*.